\documentclass[aps,twocolumn,showpacs,amsmath,superscriptaddress]{revtex4}
\usepackage{graphicx}

\begin{document}
\title{Bose-Hubbard model with attractive interactions} 
\author{Michael W. \surname{Jack}}
\affiliation{NTT Basic Research Laboratories, NTT Corporation, 3-1 Morinosato Wakamiya, Atsugi-shi, Kanagawa 243-0198, Japan}
\affiliation{Department of Physics and Astronomy and Rice Quantum Institute, Rice University, Houston, Texas 77251}
\author{Makoto \surname{Yamashita}}

\affiliation{NTT Basic Research Laboratories, NTT Corporation, 3-1 Morinosato Wakamiya, Atsugi-shi, Kanagawa 243-0198, Japan}

\date{\today}
\begin{abstract}
We consider the Bose-Hubbard model of atoms in an optical lattice potential  when the atom-atom interactions are attractive. If the lowest energy lattice sites are degenerate (such as in the homogeneous case), then, at a critical value of the interaction strength, a phase-coherent condensate becomes  unstable to a  quantum superposition such that the number distribution of each of the degenerate sites becomes double peaked. In the limit when the interaction dominates, the superposition becomes macroscopic and has the form $|\psi\rangle\propto\sum_{j} e^{i\phi_{j}}\hat{b}^{\dagger N}_{j}|{\rm vac}\rangle$, where $N$ is the total number of atoms and the sum ranges over the energy-degenerate sites.  
\end{abstract}
\maketitle 
An optical lattice potential loaded with a Bose-Einstein condensate of neutral atoms 
was predicted  by  Jaksch {\em et al.}\cite{jaksch98} and recently confirmed by Greiner {\em et al.}\cite{greiner02} to be a realization of the Bose-Hubbard model of condensed matter physics\cite{fisher89}. In addition to exemplifying a quantum phase transition, this system is ideal for creating and controlling the quantum states of the atoms at the sites of the lattice potential\cite{orzel01,greiner02}. The Bose-Hubbard model is described by a Hamiltonian of the form\cite{jaksch98}
\begin{equation}
H=-J\sum_{\langle j,i\rangle} \hat{b}_{j}^{\dagger}\hat{b}_{i}+\sum_{i}\epsilon_{i}\hat{n}_{i}+\frac{1}{2}U\hat{n}_{i}(\hat{n}_{i}-1)\label{hamiltonian}
\end{equation}
 where $\hat{b}_{i}$ and $\hat{n}_{i}=\hat{b}^{\dagger}_{i}\hat{b}_{i}$ are the annihilation and number operators of the mode localized at the $i$th lattice site. $J$ is the hopping matrix to neighboring sites,  $U\propto a_{s}$ is the strength of the on-site interactions due to $s$-wave scattering and $\epsilon_{i}$ is an energy off-set due to an additional confining magnetic trap.  
In the large hopping regime, $J\gg |U|$, the ground state of this system is well described by a phase-coherent condensate and the number fluctuations are approximately Poissonian: $\delta n_{i}\equiv\sqrt{\langle \hat{n}^{2}_{i}\rangle-n_{i}^{2}}\approx \sqrt{n_{i}}$, where $n_{i}=\langle\hat{n}_{i}\rangle$. For $U>0$, increasing the interaction energy reduces  the on-site number fluctuations, ($\delta n_{i}<\sqrt{n_{i}}$) and,  at a critical value of $U/J$, the system undergoes a Mott-insulator phase transition ($\delta n_{i}\rightarrow 0$) and all phase coherence vanishes\cite{greiner02}. In the experimental realization, the simple global phase-transition picture is complicated by the presence of the confining trap which produces local Mott domains\cite{batrouni02}.

The use of neutral atoms opens up the possibility of exploring the Bose-Hubbard model with attractive interactions $(U<0)$ as certain species of atom interact via a  negative $s$-wave scattering length and many atoms can be made to interact via a negative scattering length by using the technique of Feshbach resonance to alter the interaction potential\cite{inouye98}. Attractive interactions are particularly interesting because they lead to an instability of a phase-coherent condensate\cite{ruprecht95,baym96,yurovsky02,gerton00,donley01,chin03}. The aim of this letter is to describe the nature of this instability for the Bose-Hubbard model and it's relationship to the formation of novel quantum states.

Certain aspects of the behavior of the attractive Bose-Hubbard model can be deduced from the results for a simple two-mode model of a condensate in a double-well potential\cite{steel98,cirac98feb}. In contrast to the repulsive case, it is predicted that the number fluctuations will increase ($\delta n_{i}>\sqrt{n_{i}}$) as the magnitude of the interaction energy is increased\cite{steel98}. The behavior as the magnitude of the interaction energy is further increased depends crucially on the single-particle energies of the two wells, $\epsilon_{1}$ and $\epsilon_{2}$. In the case where the wells are asymmetric, $\epsilon_{1}\neq\epsilon_{2}$, the energy is minimized by all atoms accumulating in the lower energy site.  In the case of symmetric wells, $\epsilon_{1}=\epsilon_{2}$,
Cirac {\em et al.} \cite{cirac98feb} and Steel {\em et al.} \cite{steel98} have shown  that the system is unable to choose which site to accumulate in and will  form a quantum superposition of  the two possibilities. This superposition state is associated with very large number fluctuations: $\delta n_{i}\rightarrow n_{i}$. In the multi-well case considered here, we expect this superposition state to form between all sites that are degenerate with the lowest energy site. 

These results show that if the site energies are degenerate, then a phase-coherent condensate will become unstable to a superposition state. On the other hand, in the absence of a lattice potential, a condensate with attractive interactions is known to become unstable as the interaction strength is increased\cite{ruprecht95,baym96,yurovsky02,gerton00,donley01,chin03}. It is convenient to consider two distinct types of instability: (I) a global implosion of the condensate wavefunction confined in a harmonic trap\cite{ruprecht95,baym96} and (II) local instabilities of an {\em unconfined} condensate accompanied by large density fluctuations\cite{yurovsky02,chin03}. In the present case we find no evidence of an instability of type I, even with the addition of a harmonic confining trap, and conclude that the Bose-Hubbard model becomes invalid before this type of instability can occur. However, the instability where a condensate gives way to a superposition state described here shows  many similarities to an instability of type II. 

We can analyze the stability of a condensate in a lattice potential via a Bogoliubov type treatment\cite{oosten01,burnett02}: in the limit of large hopping, $J\gg |U|$, the ground state is well approximated by a phase-coherent condensate described by a mean-field, and we can  consider small fluctuations about this mean-field by making the replacement
$\hat{b}_{i}(t)=e^{-i\mu t/\hbar}[\beta_{i}+\hat{\delta}_{i}(t)]$ 
in the Heisenberg equations of motion for $\hat{b}_{i}$ and neglecting all terms except those linear in the fluctuations $\hat{\delta}_{i}$. This results in the  equations 
\begin{eqnarray}
0&=&-J\sum_{\langle j,i\rangle}\beta_{j}+(\epsilon_{i}-\mu+U|\beta_{i}|^2)\beta_{i}\label{bogoliubovEqns1}\\
\frac{d\hat{\delta}_{i}}{dt} &=&-J\sum_{\langle j,i\rangle}\hat{\delta}_{j}+(2U|\beta_{i}|^{2}+\epsilon_{i}-\mu)\hat{\delta}_{i}+U\beta_{i}^{2}\hat{\delta}^{\dagger}_{i}\label{bogoliubovEqns2}
\end{eqnarray}
for the mean field and fluctuations, respectively. After solving Eq.(\ref{bogoliubovEqns1}) for $\beta_{i}$ and $\mu$, Eq.(\ref{bogoliubovEqns2}) can be solved by making the Bogoliubov transformation $\hat{\delta}_{i}=\sum_{k}u_{i,k}e^{-i\omega_{k}t}\hat{\delta}_{k}+v_{i,k}^{*}e^{i\omega_{k}t}\hat{\delta}_{{k}}^{\dagger}$, with the normalization condition $\sum_{i} |u_{i,{k}}|^{2}-|v_{i,{k}}|^{2}=1$, and solving the resulting equations for $\omega_{{k}}$, $u_{i,{k}}$ and $v_{i,{k}}$.

Assuming a homogeneous one-dimensional lattice of $M$ sites (with periodic boundary conditions) containing $N$ atoms, equations (\ref{bogoliubovEqns1}) and (\ref{bogoliubovEqns2}) can be solved analytically \cite{oosten01,burnett02} such that $\mu=Un-2J$, where $n=|\beta_{i}|^{2}=N/M$ is the mean-field solution and the quasi-particle energies  have the form 
\begin{equation}
\hbar\omega_{k}=\sqrt{\varepsilon_{k}(\varepsilon_{k}+2 U n)}\label{quasiparticleFrequencies}
\end{equation}
where $\varepsilon_{k}=4J\sin^{2}(a k/2)$. Here $k=(2\pi/aM) m$ for $m=-M/2,\cdots,M/2$. This shows that as $U$ becomes increasingly negative, the first quasi-particle energy (coresponding to $k=2\pi/aM$) drops to zero at $Un=-2J\sin^{2}(\pi/M)$ and then becomes imaginary, signaling a critical point beyond which the lattice system is unable to support a condensate. An analytical  expression for the on-site number fluctuations can also be determined from this treatment \cite{burnett02} (see also \cite{javanainen99dec}) as: 
\begin{equation}
\delta n_{i}^{2} =\frac{n}{M}\sum_{k}\frac{\varepsilon_{k}}{\hbar \omega_{k}}\label{numberFluctuations}.
\end{equation}
 It is evident that at the critical point the number fluctuations (\ref{numberFluctuations})  diverge.  Comparing with Ref.\cite{yurovsky02} we see that a type II instability is formally very similar to the present case.

\begin{figure}
  \begin{center}
    \includegraphics[scale=0.42]{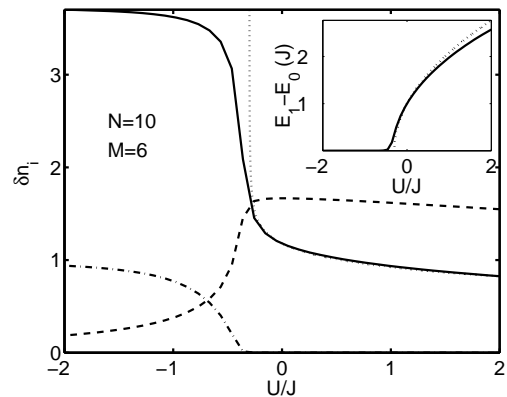}
    \caption{\protect\label{homogeneous} Ground state of a homogeneous one-dimensional lattice. This figure shows $\delta n_{i}$ (solid line), $C_{1}$ (dashed line) and $C_{N}$ (dash-dotted line) as a function of $U/J$. The inset shows the energy difference between the first excited state and the ground state. The dotted lines correspond to a  Bogoliubov treatment. }
  \end{center}
\end{figure}

In order to treat  the strong interaction regime beyond the instability (where the Bogoliubov treatment breaks down), we have numerically calculated the exact ground state. In the Fock state basis the state space of the system is large: $(N+M-1)!/[N!(M-1)!]$. But the Hamiltonian (\ref{hamiltonian}) is a very sparse matrix and so for small numbers of atoms and sites we can calculate the lowest few eigenvalues and eigenstates by the Lanczos method\cite{lanchos}. The results of these calculations are  presented in Fig.\ref{homogeneous}. 
After the critical point of the Bogoliubov treatment the exact calculations show that the number distribution becomes double peaked 
(corresponding to the formation of a superposition state) which gives rise to the sudden increase in the number fluctuations shown in Fig.\ref{homogeneous}. The two peaks of the distribution move further apart and narrow as the interaction is increased which reduces the single-particle correlation between neighboring sites, $C_{1}=\langle \hat{b}_{i}^{\dagger}\hat{b}_{i+1}\rangle$, but increases the $N$-particle correlation, $C_{N}=M\langle \hat{b}_{i}^{\dagger N}\hat{b}_{i+1}^{N}\rangle/N!$. Finally, in the strong attractive interaction limit, the results confirm that the ground state is a macroscopic superposition of the form: 
\begin{equation}
|\psi\rangle=\frac{1}{\sqrt{MN!}}\sum_{j=1}^{M}e^{i\phi_{j}}\hat{b}^{\dagger N}_{j}|{\rm vac}\rangle\label{superposition},
\end{equation}
and the number fluctuations become $\delta n_{i}=N\sqrt{M-1}/M$ or $\approx N/\sqrt{M}$ for many sites.

\begin{figure}
  \begin{center}
\includegraphics[scale=0.45]{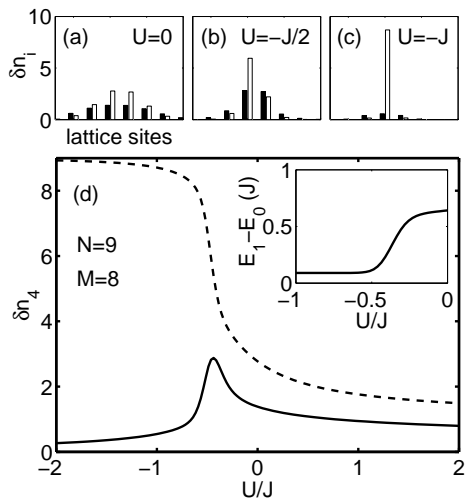}
                         \caption{\protect\label{inhomogeneous}Ground state in the presence of a harmonic confining trap with $\lambda=0.1J$ and $\Delta=0.9$. Figures (a)-(c) show $n_{i}$ (white bars) and $\delta n_{i}$ (black bars) at each site for various values of $U/J$. Figure (d) shows $n_{4}$ (dashed line) and   $\delta n_{4}$ (solid line)  as a function of $U/J$.  
The inset in (d) shows the energy difference between the first excited state and the ground state.
}
  \end{center}
\end{figure}

In the current experiments\cite{greiner02,orzel01}, an inhomogeneity is introduced to the lattice system by an harmonic magnetic trap which is used to confine the atoms in space. In the one-dimensional case considered here, this gives rise to the single-particle energies $\epsilon_{i}=\lambda[i-\frac{1}{2}(M+\Delta)]^{2}$, where $i=1$ to $M$, $\lambda$ is a measure of the curvature and $0\leq \Delta<1$ is the offset of the lattice from the center of the confining potential. Figure \ref{inhomogeneous} shows the results of exact calculations  in the case of non-degenerate sites. Note that the number fluctuations become large close to where the critical point would have occurred if the sites were degenerate (the system gets close to forming a superposition at this point) and then decrease as the  atoms accumulate into a single site. 

\begin{figure}
\begin{center}
\includegraphics[scale=0.35]{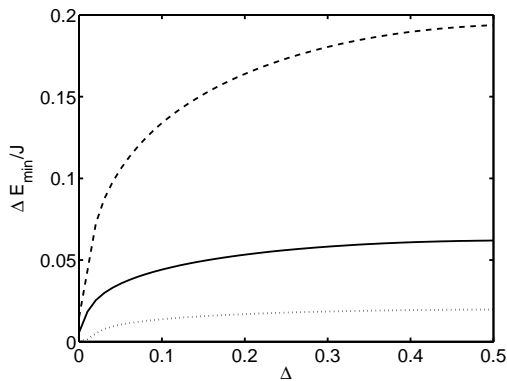}
\caption{\protect\label{gap} Minimum energy gap between the ground state and first excited state as a function of $\Delta$ for $M=50$ sites. The values of the curvature are $\lambda=10^{-3}$ (solid line), $\lambda=10^{-2}$ (dashed line) and $\lambda=10^{-4}$ (dotted line). }
\end{center}
\end{figure}

We can also calculate the critical behavior of the inhomogeneous system via a Bogoliubov treatment, which requires the numerical solution of the mean-field equations (\ref{bogoliubovEqns1}) before solving the linear equations (\ref{bogoliubovEqns2}) for the quasi-particles.   In the non-degenerate case when $0<\Delta<1$, for $M$ odd, there is always one site, $i=\frac{1}{2}(M+1)$, with the lowest single-particle energy. In this case, no critical point is seen and as the interaction becomes more attractive, the condensate--which has an approximately Gaussian spatial profile of width $w$--simply decreases in width until all atoms accumulate in site $i=\frac{1}{2}(M+1)$. 
In figure \ref{gap} we have plotted the minimum energy gap to the first excited state, $\Delta E_{\rm min}$, (found by varying $U/J$ over a broad range of values) as a function of $\Delta$.
The finite value of this gap for $\Delta\neq 0$ demonstrates the stability of the condensate in the non-degenerate case. We find no evidence of an instability of type I which would be expected to occur even in the non-degenerate case. 

\begin{figure}
\begin{center}
 \includegraphics[scale=0.35]{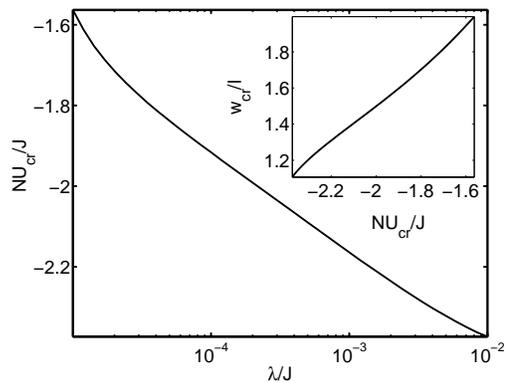}
\caption{\protect\label{CP} Plot of the critical value of $U/J$ as a function of the curvature of the confining trap, $\lambda$, in the case of two degenerate sites at the center of the trap. The inset shows the width $w$ of the condensate at the critical point.}
\end{center}
\end{figure}

Figure \ref{CP} shows the dependence of the critical point (where the first excitation energy becomes imaginary) on the curvature in the degenerate case: $\Delta=0$.
The inset shows that for a broad range of trap curvatures, the critical point occurs when the condensate width is of the order of the site spacing.

\begin{figure}
  \begin{center}
    \includegraphics[scale=0.4]{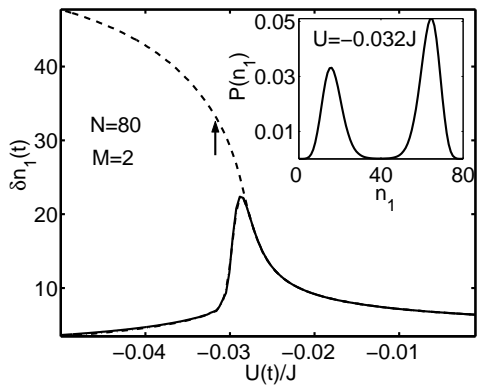}
    \caption{\protect\label{nonadiabatic} Non-adiabatic evolution of a two-mode model from the ground state at $U(t=0)=0$ with the parameters $U_{f}=-0.05J$, $\epsilon_{2}-\epsilon_{1}=0.0005J$. This figure shows the time evolution of the number fluctuations, $\delta n_{1}$  for the rates $\gamma<0.005J/\hbar$ (solid line) and $\gamma=0.01J/\hbar$ (dashed line).  The inset shows the two-peaked  structure (coresponding to a superposition state) of the number distribution, $P(n_{1})=|\langle \psi(t)|n_{1},N-n_{1}\rangle|^{2}$, at the point corresponding to the arrow.}
  \end{center}
\end{figure}

In a realistic situation it may be difficult (if not impossible) to create a lattice with exactly degenerate sites.  However, in current experiments, the high interaction regime is reached by a {\em dynamic} process whereby the relevant parameter, $\kappa(t)=|U(t)|/J(t)$, is increased at a certain rate $\gamma$ (this is normally in the adiabatic regime so the system remains in the ground state). In this case, close to $\kappa(t)\sim 1/N$ (where the critical point would be if the sites were degenerate), if one increases $\kappa(t)$ at a rate faster than the oscillation frequency between two different sites, $\gamma\gg (\epsilon_{1}-\epsilon_{2})/\hbar$, (but still slower than the tunneling rate) then, for short times, the system will be unaware of the inhomogeneity and it is possible to effectively ``capture'' the superposition before the atoms are able to tunnel into the lowest energy site. To illustrate this point we have numerically solved the Schr\"{o}dinger equation for a two-mode model with the time-dependent interaction strength $U(t)=U_{f}(1-e^{-\gamma t})$. The results of this simulation are shown in Fig.\ref{nonadiabatic}. The large number fluctuations  for a sufficiently large $\gamma$ (dotted line) confirm the formation of a superposition even though the single-particle energies in the two wells differ.  

Experimental realization of the interaction dominated regime of the Bose-Hubbard model with a large number of atoms is complicated by the fact that the localization of all $N$ atoms at a single lattice site may render the Bose-Hubbard model invalid unless the magnitude of the scattering length is small. Approximating the potential at each lattice site by a harmonic potential of length $a_{\rm ho}$, the interaction strength must satisfy $n_{i}a_{\rm ho}\gg\langle \hat{n}_{i}(\hat{n}_{i}-1)\rangle |a_{s}|$, such that the interaction does not alter the shape of the localized mode-functions at each site [In fact, we can continue to use the Bose-Hubbard model beyond this inequality with renormalized parameters\cite{vanoosten02}, the ultimate limit being the stability of the localized mode-function against a type I instability: $a_{\rm ho}>n_{i}|a_{s}|$].  A possible method to overcome this is to load a very deep optical lattice (so that $J$ is small) with repulsively interacting atoms and use Feshbach resonance to slowly tune the interaction through zero so it becomes just slightly negative, as described in Ref.\cite{strecker02} for $^{7}$Li.
This method has the additional advantage that it will minimize three-body loss of atoms which scales as $a_{s}^{4}$\cite{esry99}.

In conclusion, the ground state of the attractive Bose-Hubbard model displays behavior fundamentally different from the repulsive case. In particular, if the lowest energy sites are degenerate then, at a critical value of the interaction strength, a phase-coherent condensate becomes  unstable to a quantum superposition such that the number distribution at each degenerate site becomes double peaked. The atoms have a tendancy to accumulate at a single site in order to minimize the interaction energy but they are unable to choose which site due to the energy degeneracy and so form a superposition of all the possibilities. Interestingly, our results suggest  that the Bose-Hubbard model becomes invalid before an instability of type I (for a confined condensate) occurs and that an instability of type II (for an unconfined condensate) corresponds to the formation of superposition states due to the homogeneity of free space.  In an  experimental realization, atom loss (or absorption imaging) will destroy the superposition by tending to localize the atoms at one site.  Superpositions such as (\ref{superposition}) can be destroyed by the loss of just one atom which will ``collapse'' the quantum state of the atoms to one of the degenerate lattice sites. Less macroscopic superpositions, such as those formed just after the critical point, will be more robust against loss\cite{cirac98feb}. Methods of non-destructive detection of these superposition states will be the topic of future work.


The authors would like to thank A. Kawaguchi, Y. Tokura and R. Hulet for useful discussions.

\begin{thebibliography}{21}
\expandafter\ifx\csname natexlab\endcsname\relax\def\natexlab#1{#1}\fi
\expandafter\ifx\csname bibnamefont\endcsname\relax
  \def\bibnamefont#1{#1}\fi
\expandafter\ifx\csname bibfnamefont\endcsname\relax
  \def\bibfnamefont#1{#1}\fi
\expandafter\ifx\csname citenamefont\endcsname\relax
  \def\citenamefont#1{#1}\fi
\expandafter\ifx\csname url\endcsname\relax
  \def\url#1{\texttt{#1}}\fi
\expandafter\ifx\csname urlprefix\endcsname\relax\def\urlprefix{URL }\fi
\providecommand{\bibinfo}[2]{#2}
\providecommand{\eprint}[2][]{\url{#2}}

\bibitem[{\citenamefont{Jaksch et~al.}(1998)\citenamefont{Jaksch, Bruder,
  Cirac, Gardiner, and Zoller}}]{jaksch98}
\bibinfo{author}{\bibfnamefont{D.}~\bibnamefont{Jaksch}},
  \bibinfo{author}{\bibfnamefont{C.}~\bibnamefont{Bruder}},
  \bibinfo{author}{\bibfnamefont{J.~I.} \bibnamefont{Cirac}},
  \bibinfo{author}{\bibfnamefont{C.~W.} \bibnamefont{Gardiner}},
  \bibnamefont{and} \bibinfo{author}{\bibfnamefont{P.}~\bibnamefont{Zoller}},
  \bibinfo{journal}{Phys. Rev. Lett.} \textbf{\bibinfo{volume}{81}},
  \bibinfo{pages}{3108} (\bibinfo{year}{1998}).

\bibitem[{\citenamefont{Greiner et~al.}(2002)\citenamefont{Greiner, Mandel,
  Esslinger, H\"{a}nsch, and Bloch}}]{greiner02}
\bibinfo{author}{\bibfnamefont{M.}~\bibnamefont{Greiner}},
  \bibinfo{author}{\bibfnamefont{O.}~\bibnamefont{Mandel}},
  \bibinfo{author}{\bibfnamefont{T.}~\bibnamefont{Esslinger}},
  \bibinfo{author}{\bibfnamefont{T.~W.} \bibnamefont{H\"{a}nsch}},
  \bibnamefont{and} \bibinfo{author}{\bibfnamefont{I.}~\bibnamefont{Bloch}},
  \bibinfo{journal}{Nature (London)} \textbf{\bibinfo{volume}{415}},
  \bibinfo{pages}{39} (\bibinfo{year}{2002}).

\bibitem[{\citenamefont{Fisher et~al.}(1989)\citenamefont{Fisher, Weichman,
  Grinstein, and Fisher}}]{fisher89}
\bibinfo{author}{\bibfnamefont{M.~P.~A.} \bibnamefont{Fisher}},
  \bibinfo{author}{\bibfnamefont{P.~B.} \bibnamefont{Weichman}},
  \bibinfo{author}{\bibfnamefont{G.}~\bibnamefont{Grinstein}},
  \bibnamefont{and} \bibinfo{author}{\bibfnamefont{D.~S.}
  \bibnamefont{Fisher}}, \bibinfo{journal}{Phys. Rev. B}
  \textbf{\bibinfo{volume}{40}}, \bibinfo{pages}{546} (\bibinfo{year}{1989}).

\bibitem[{\citenamefont{Orzel et~al.}(2001)\citenamefont{Orzel, Tuchman,
  Fenselau, Yasuda, and Kasevich}}]{orzel01}
\bibinfo{author}{\bibfnamefont{C.}~\bibnamefont{Orzel}},
  \bibinfo{author}{\bibfnamefont{A.~K.} \bibnamefont{Tuchman}},
  \bibinfo{author}{\bibfnamefont{M.~L.} \bibnamefont{Fenselau}},
  \bibinfo{author}{\bibfnamefont{M.}~\bibnamefont{Yasuda}}, \bibnamefont{and}
  \bibinfo{author}{\bibfnamefont{M.~A.} \bibnamefont{Kasevich}},
  \bibinfo{journal}{Science} \textbf{\bibinfo{volume}{291}},
  \bibinfo{pages}{2386} (\bibinfo{year}{2001}).

\bibitem[{\citenamefont{Batrouni et~al.}(2002)\citenamefont{Batrouni, Rousseau,
  Scalettar, Rigol, Muramatsu, Denteneer, and Troyer}}]{batrouni02}
\bibinfo{author}{\bibfnamefont{G.~G.} \bibnamefont{Batrouni}},
  \bibinfo{author}{\bibfnamefont{V.}~\bibnamefont{Rousseau}},
  \bibinfo{author}{\bibfnamefont{R.~T.} \bibnamefont{Scalettar}},
  \bibinfo{author}{\bibfnamefont{M.}~\bibnamefont{Rigol}},
  \bibinfo{author}{\bibfnamefont{A.}~\bibnamefont{Muramatsu}},
  \bibinfo{author}{\bibfnamefont{P.~J.~H.} \bibnamefont{Denteneer}},
  \bibnamefont{and} \bibinfo{author}{\bibfnamefont{M.}~\bibnamefont{Troyer}},
  \bibinfo{journal}{Phys. Rev. Lett.} \textbf{\bibinfo{volume}{89}},
  \bibinfo{pages}{117203} (\bibinfo{year}{2002}).

\bibitem[{\citenamefont{Inouye et~al.}(1998)\citenamefont{Inouye, Mathews,
  Stenger, Miesner, Stamper-Kurn, and Ketterle}}]{inouye98}
\bibinfo{author}{\bibfnamefont{S.}~\bibnamefont{Inouye}},
  \bibinfo{author}{\bibfnamefont{M.~R.} \bibnamefont{Mathews}},
  \bibinfo{author}{\bibfnamefont{J.}~\bibnamefont{Stenger}},
  \bibinfo{author}{\bibfnamefont{H.~J.} \bibnamefont{Miesner}},
  \bibinfo{author}{\bibfnamefont{D.~M.} \bibnamefont{Stamper-Kurn}},
  \bibnamefont{and} \bibinfo{author}{\bibfnamefont{W.}~\bibnamefont{Ketterle}},
  \bibinfo{journal}{Nature (London)} \textbf{\bibinfo{volume}{392}},
  \bibinfo{pages}{151} (\bibinfo{year}{1998}).

\bibitem[{\citenamefont{Ruprecht et~al.}(1995)\citenamefont{Ruprecht, Holland,
  and Burnett}}]{ruprecht95}
\bibinfo{author}{\bibfnamefont{P.~A.} \bibnamefont{Ruprecht}},
  \bibinfo{author}{\bibfnamefont{M.~J.} \bibnamefont{Holland}},
   \bibinfo{author}{\bibfnamefont{K.}~\bibnamefont{Burnett}} \bibnamefont{and}, \bibinfo{author}{\bibfnamefont{M.}~\bibnamefont{Edwards}}
  \bibinfo{journal}{Phys. Rev. A} \textbf{\bibinfo{volume}{51}},
  \bibinfo{pages}{4704} (\bibinfo{year}{1995}).

\bibitem[{\citenamefont{Baym and Pethick}(1996)}]{baym96}
\bibinfo{author}{\bibfnamefont{G.}~\bibnamefont{Baym}} \bibnamefont{and}
  \bibinfo{author}{\bibfnamefont{C.~J.} \bibnamefont{Pethick}},
  \bibinfo{journal}{Phys. Rev. Lett.} \textbf{\bibinfo{volume}{76}},
  \bibinfo{pages}{6} (\bibinfo{year}{1996}).

\bibitem[{\citenamefont{Yurovsky}(2002)}]{yurovsky02}
\bibinfo{author}{\bibfnamefont{V.~A.} \bibnamefont{Yurovsky}},
  \bibinfo{journal}{Phys. Rev. A} \textbf{\bibinfo{volume}{65}},
  \bibinfo{pages}{033605} (\bibinfo{year}{2002}).

\bibitem[{\citenamefont{Gerton et~al.}(2000)\citenamefont{Gerton, Strekalov,
  Prodan, and Hulet}}]{gerton00}
\bibinfo{author}{\bibfnamefont{J.~M.} \bibnamefont{Gerton}},
  \bibinfo{author}{\bibfnamefont{D.}~\bibnamefont{Strekalov}},
  \bibinfo{author}{\bibfnamefont{I.}~\bibnamefont{Prodan}}, \bibnamefont{and}
  \bibinfo{author}{\bibfnamefont{R.~G.} \bibnamefont{Hulet}},
  \bibinfo{journal}{Nature (London)} \textbf{\bibinfo{volume}{408}},
  \bibinfo{pages}{692} (\bibinfo{year}{2000}).

\bibitem[{\citenamefont{Donley et~al.}(2001)\citenamefont{Donley, Claussen,
  Cornish, Roberts, Cornell, and Wienman}}]{donley01}
\bibinfo{author}{\bibfnamefont{E.~A.} \bibnamefont{Donley}},
  \bibinfo{author}{\bibfnamefont{N.~R.} \bibnamefont{Claussen}},
  \bibinfo{author}{\bibfnamefont{S.~L.} \bibnamefont{Cornish}},
  \bibinfo{author}{\bibfnamefont{J.~L.} \bibnamefont{Roberts}},
  \bibinfo{author}{\bibfnamefont{E.~A.} \bibnamefont{Cornell}},
  \bibnamefont{and} \bibinfo{author}{\bibfnamefont{C.~E.}
  \bibnamefont{Wienman}}, \bibinfo{journal}{Nature (London)}
  \textbf{\bibinfo{volume}{412}}, \bibinfo{pages}{295} (\bibinfo{year}{2001}).

\bibitem[{\citenamefont{Chin et~al.}(2003)\citenamefont{Chin, Vogels, and
  Ketterle}}]{chin03}
\bibinfo{author}{\bibfnamefont{J.~K.} \bibnamefont{Chin}},
  \bibinfo{author}{\bibfnamefont{J.~M.} \bibnamefont{Vogels}},
  \bibnamefont{and} \bibinfo{author}{\bibfnamefont{W.}~\bibnamefont{Ketterle}},
  \bibinfo{journal}{Phys. Rev. Lett.} \textbf{\bibinfo{volume}{90}},
  \bibinfo{pages}{160405} (\bibinfo{year}{2003}).

\bibitem[{\citenamefont{Steel and Collett}(1998)}]{steel98}
\bibinfo{author}{\bibfnamefont{M.~J.} \bibnamefont{Steel}} \bibnamefont{and}
  \bibinfo{author}{\bibfnamefont{M.~J.} \bibnamefont{Collett}},
  \bibinfo{journal}{Phys. Rev. A} \textbf{\bibinfo{volume}{57}},
  \bibinfo{pages}{2920} (\bibinfo{year}{1998}).

\bibitem[{\citenamefont{Cirac et~al.}(1998)\citenamefont{Cirac, Lewenstein,
  M\/{o}lmer, and Zoller}}]{cirac98feb}
\bibinfo{author}{\bibfnamefont{J.~I.} \bibnamefont{Cirac}},
  \bibinfo{author}{\bibfnamefont{M.}~\bibnamefont{Lewenstein}},
  \bibinfo{author}{\bibfnamefont{K.}~\bibnamefont{M\/{o}lmer}},
  \bibnamefont{and} \bibinfo{author}{\bibfnamefont{P.}~\bibnamefont{Zoller}},
  \bibinfo{journal}{Phys. Rev. A} \textbf{\bibinfo{volume}{57}},
  \bibinfo{pages}{1208} (\bibinfo{year}{1998}).

\bibitem[{\citenamefont{van Oosten et~al.}(2001)\citenamefont{van Oosten,
  van~der Straten, and Stoof}}]{oosten01}
\bibinfo{author}{\bibfnamefont{D.}~\bibnamefont{van Oosten}},
  \bibinfo{author}{\bibfnamefont{P.}~\bibnamefont{van~der Straten}},
  \bibnamefont{and} \bibinfo{author}{\bibfnamefont{H.~T.~C.}
  \bibnamefont{Stoof}}, \bibinfo{journal}{Phys. Rev. A}
  \textbf{\bibinfo{volume}{63}}, \bibinfo{pages}{053601}
  (\bibinfo{year}{2001}).

\bibitem[{\citenamefont{Burnett et~al.}(2002)\citenamefont{Burnett, Edwards,
  Clark, and Shotter}}]{burnett02}
\bibinfo{author}{\bibfnamefont{K.}~\bibnamefont{Burnett}},
  \bibinfo{author}{\bibfnamefont{M.}~\bibnamefont{Edwards}},
  \bibinfo{author}{\bibfnamefont{C.~W.} \bibnamefont{Clark}}, \bibnamefont{and}
  \bibinfo{author}{\bibfnamefont{M.}~\bibnamefont{Shotter}},
  \bibinfo{journal}{J. Phys. B: At. Mol. Opt. Phys.}
  \textbf{\bibinfo{volume}{35}}, \bibinfo{pages}{1671} (\bibinfo{year}{2002}).

\bibitem[{\citenamefont{Javanainen}(1999)}]{javanainen99dec}
\bibinfo{author}{\bibfnamefont{J.}~\bibnamefont{Javanainen}},
  \bibinfo{journal}{Phys. Rev. A} \textbf{\bibinfo{volume}{60}},
  \bibinfo{pages}{4902} (\bibinfo{year}{1999}).

\bibitem[{\citenamefont{Bai et~al.}(2000)\citenamefont{Bai, Demmel, Dongarra,
  Ruhe, and {H. van der Vorst}}}]{lanchos}
\bibinfo{editor}{\bibfnamefont{Z.}~\bibnamefont{Bai}},
  \bibinfo{editor}{\bibfnamefont{J.}~\bibnamefont{Demmel}},
  \bibinfo{editor}{\bibfnamefont{J.}~\bibnamefont{Dongarra}},
  \bibinfo{editor}{\bibfnamefont{A.}~\bibnamefont{Ruhe}}, \bibnamefont{and}
  \bibinfo{editor}{\bibnamefont{{H. van der Vorst}}}, eds.,
  \emph{\bibinfo{title}{Templates for the Solution of Algebraic Eigenvalue
  Problems: A Practical Guide.}} (\bibinfo{publisher}{SIAM},
  \bibinfo{year}{2000}).

\bibitem[{\citenamefont{van Oosten et~al.}(2003)\citenamefont{van Oosten,
  van~der Straten, and Stoof}}]{vanoosten02}
\bibinfo{author}{\bibfnamefont{D.}~\bibnamefont{van Oosten}},
  \bibinfo{author}{\bibfnamefont{P.}~\bibnamefont{van~der Straten}},
  \bibnamefont{and} \bibinfo{author}{\bibfnamefont{H.~T.~C.}
  \bibnamefont{Stoof}}, \bibinfo{journal}{Phys. Rev. A}
  \textbf{\bibinfo{volume}{67}}, \bibinfo{pages}{033606}
  (\bibinfo{year}{2003}).

\bibitem[{\citenamefont{Strecker et~al.}(2002)\citenamefont{Strecker,
  Partridge, Truscott, and Hulet}}]{strecker02}
\bibinfo{author}{\bibfnamefont{K.~E.} \bibnamefont{Strecker}},
  \bibinfo{author}{\bibfnamefont{G.~B.} \bibnamefont{Partridge}},
  \bibinfo{author}{\bibfnamefont{A.~G.} \bibnamefont{Truscott}},
  \bibnamefont{and} \bibinfo{author}{\bibfnamefont{R.~G.} \bibnamefont{Hulet}},
  \bibinfo{journal}{Nature (London)} \textbf{\bibinfo{volume}{417}},
  \bibinfo{pages}{150} (\bibinfo{year}{2002}).

\bibitem[{\citenamefont{Esry et~al.}(1999)\citenamefont{Esry, Greene, and
  Burke}}]{esry99}
\bibinfo{author}{\bibfnamefont{B.~D.} \bibnamefont{Esry}},
  \bibinfo{author}{\bibfnamefont{C.~H.} \bibnamefont{Greene}},
  \bibnamefont{and} \bibinfo{author}{\bibfnamefont{J.~P.} \bibnamefont{Burke}},
  \bibinfo{journal}{Phys. Rev. Lett.} \textbf{\bibinfo{volume}{83}},
  \bibinfo{pages}{1751} (\bibinfo{year}{1999}).

\end{thebibliography}

\end{document}